\documentclass[conference]{IEEEtran}
\usepackage{amsmath,amssymb,amsfonts}
\usepackage{graphicx}
\usepackage{algpseudocode}
\usepackage{textcomp}
\usepackage{xcolor}
\usepackage{pifont}
\usepackage{multirow}
\usepackage{blkarray} 
\usepackage{hyperref}
\usepackage{cite}
\usepackage{soul}
\usepackage[ruled,vlined]{algorithm2e} 
\usepackage{capt-of}
\usepackage{caption}
\usepackage{subcaption}
\usepackage{tikz}

\newcommand{\circleNumber}[1]{%
  \begin{tikzpicture}[baseline=(char.base)]
    \node[shape=circle, fill=black, inner sep=2pt] (char) {\textcolor{white}{#1}};
  \end{tikzpicture}
}

\SetKwInput{KwInput}{Input}                
\SetKwInput{KwOutput}{Output}              

\def\BibTeX{{\rm B\kern-.05em{\sc i\kern-.025em b}\kern-.08em
    T\kern-.1667em\lower.7ex\hbox{E}\kern-.125emX}}
    
\begin{document}
\title{\huge LIPSTICK: Corruptibi\ul{li}ty-Aware and Ex\ul{p}lainable Graph Neural Network-based Oracle-Les\ul{s} A\ul{t}tack on Log\ul{i}c Lo\ul{ck}ing}

\author{
\IEEEauthorblockN{Yeganeh Aghamohammadi}
\IEEEauthorblockA{
\textit{University of California, Santa Barbara}\\
Santa Barbara, CA, USA \\
yeganeh@ucsb.edu} 
\and
\IEEEauthorblockN{Amin Rezaei}
\IEEEauthorblockA{
\textit{California State University, Long Beach}\\
Long Beach, CA, USA \\
amin.rezaei@csulb.edu} 
}

\maketitle 

\begin{abstract} 
In a zero-trust fabless paradigm, designers are increasingly concerned about hardware-based attacks on the semiconductor supply chain. Logic locking is a design-for-trust method that adds extra key-controlled gates in the circuits to prevent hardware intellectual property theft and overproduction. While attackers have traditionally relied on an oracle to attack logic-locked circuits, machine learning attacks have shown the ability to retrieve the secret key even without access to an oracle. In this paper, we first examine the limitations of state-of-the-art machine learning attacks and argue that the use of key hamming distance as the sole model-guiding structural metric is not always useful. Then, we develop, train, and test a corruptibility-aware graph neural network-based oracle-less attack on logic locking that takes into consideration both the structure and the behavior of the circuits. Our model is explainable in the sense that we analyze what the machine learning model has interpreted in the training process and how it can perform a successful attack. Chip designers may find this information beneficial in securing their designs while avoiding incremental fixes.
\end{abstract} 

\begin{IEEEkeywords}
Logic Locking, Logic Encryption, Machine Learning, Graph Neural Networks, Corruptibility, Explainability
\end{IEEEkeywords}

\section{Introduction}\label{Sec:Intro}
While outsource manufacturing is becoming the norm in the semiconductor industry, it comes with increasing threats such as hardware Intellectual Property (IP) theft and overproduction. Logic locking \cite{Koushanfar:Logic-locking, Rajendran:toc13xor, Baumgarten:dtc10lut, Rajendran:SAR-Lock, Srivastava:Anti-SAT, Rezaei:BLE, Sinanoglu:UNSAIL, Rezaei:Cyclic, Rezaei:DLE, Rezaei:DK-Lock, Rezaei:PUF, Rezaei:Memristor, Rezaei:Sequential, Rezaei:CoLA} (a.k.a. logic encryption or logic obfuscation) is a technique to safeguard against these attacks by adding extra key-controlled gates to the circuits.

One of the well-studied threat models is the Oracle-Guided (OG) type of attacks \cite{Subramanyan:SAT, Jin:AppSAT, Shen:SigAttack, Zuzak:CLAP, Rezaei:BreakUnroll, Wang:CyclicAttack}, which assume having access to an activated Integrated Circuit (IC) purchased off the shelf in addition to a logic-locked netlist leaked from an untrusted foundry. With the aid of a Boolean satisfiability (SAT) solver, OG attacks try to prune out subsets of wrong keys by checking a relatively small number of input patterns. While the semiconductor industry must yet come up with protective mechanisms against OG attacks, Oracle-Less (OL) attacks can be detrimentally pervasive in the sense that they can leak confidential information to attackers with limited resources. Basically, we believe the weaker and yet more effective the attacker model is, the more it can be ubiquitous and harmful to hardware IP owners, and thus more critical to safeguard.

Recent advancements in Machine Learning (ML) have made it possible to propose OL attacks to predict the correct key of logic-locked circuits \cite{Sisejkovic:Snapshot, Bhunia:SAIL, Shamsi:CutSAIL, Sinanoglu:OMLA, Sinanoglu:Untangle, Sinanoglu:GNNUnlock+}. One of the suitable ML models for logic circuits is Graph Neural Network (GNN) which accepts inputs in the form of graphs and handles non-Euclidean data. GNNs are capable of learning the connections between diverse nodes and edges in a network \cite{Yu:GNN-Survey} and thus recognizing patterns in graph-structured data, such as the schematic of a logic circuit. In state-of-the-art ML-based OL attacks, the success rate is reported in terms of ML model prediction accuracy, which is defined as minimizing the hamming distance between the correct key and the key reported by the attack. 

\subsection{Research Gap}\label{Sec:Gap}
The first observation is that ML-based OL attacks are inherently \textit{approximate} attacks because of the fact that they try to find out an approximately correct key with respect to some parameters, for example, by reducing the hamming distance between the correct and reported keys. The second finding is that current OL attacks do not consider the behavior of the circuits under the reported key compared with the intended functionality, necessitating the use of more meaningful metrics, such as \textit {key precision}, which takes into account output corruptibility in addition to circuit structure. The third observation is that a holistic security assessment of logic-locking techniques is overlooked, yet \textit{explainable} ML models \cite{Ji:Explainability}, which provide us with more reliable and trustworthy predictions, need to be taken into consideration.

\begin{figure*}
\centering
\begin{minipage}{0.40\textwidth}
\centering
\includegraphics[width=0.55\textwidth]{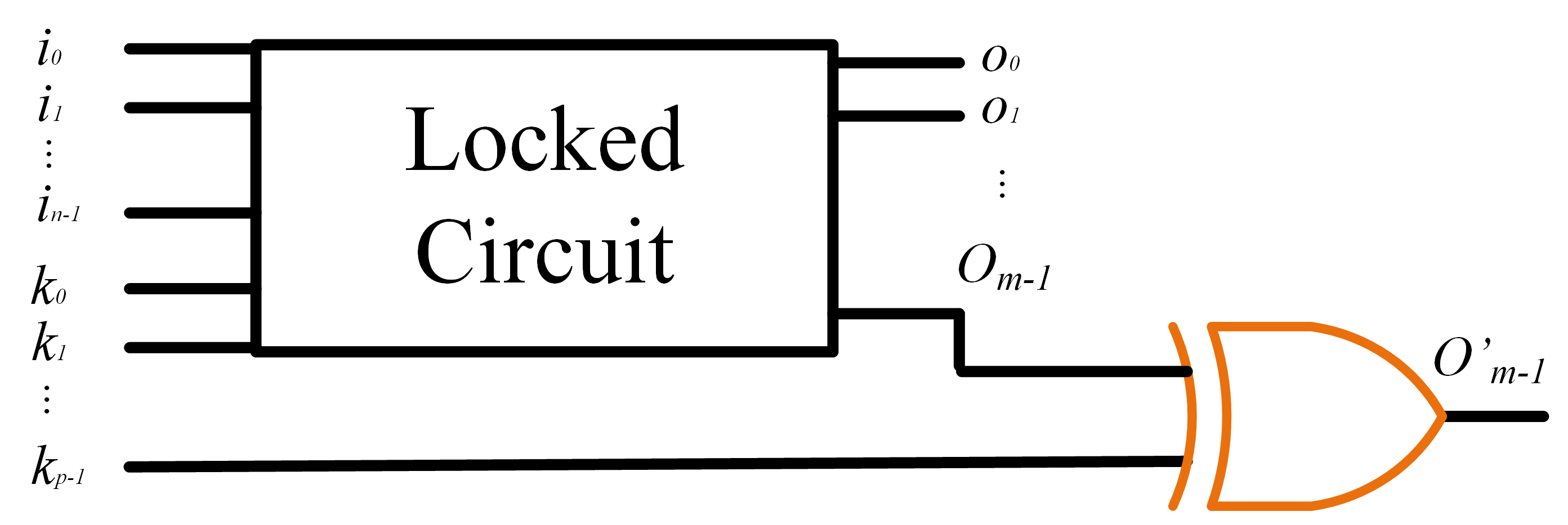}
\captionof{figure}{Counterexample for hamming distance as a key precision metric}
\label{fig:counterexample1}
\end{minipage}%
\hspace{0.5em}
\begin{minipage}{0.57\textwidth}
\centering
\includegraphics[width=0.31\textwidth]{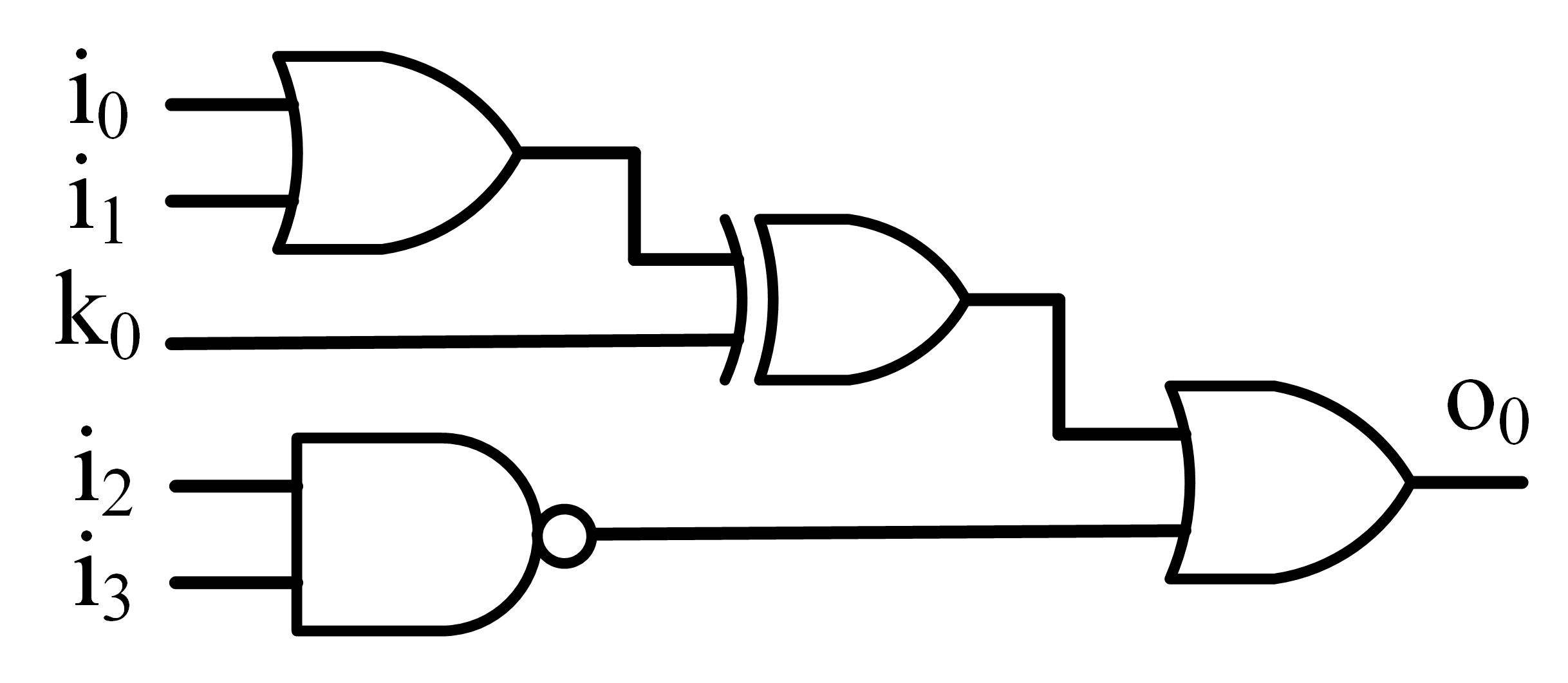}
\hspace{0.1em}
\includegraphics[width=0.31\textwidth]{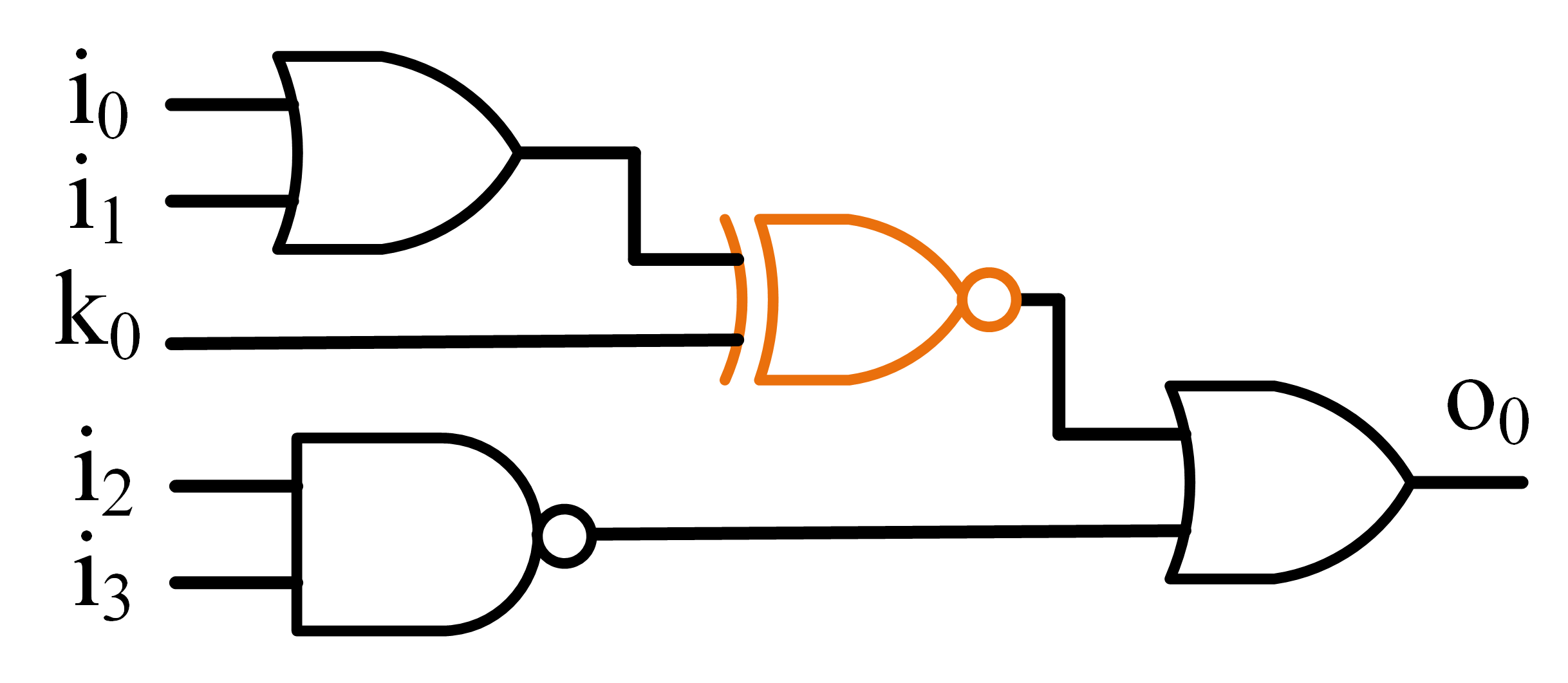} \hspace{0.1em}
\includegraphics[width=0.31\textwidth]{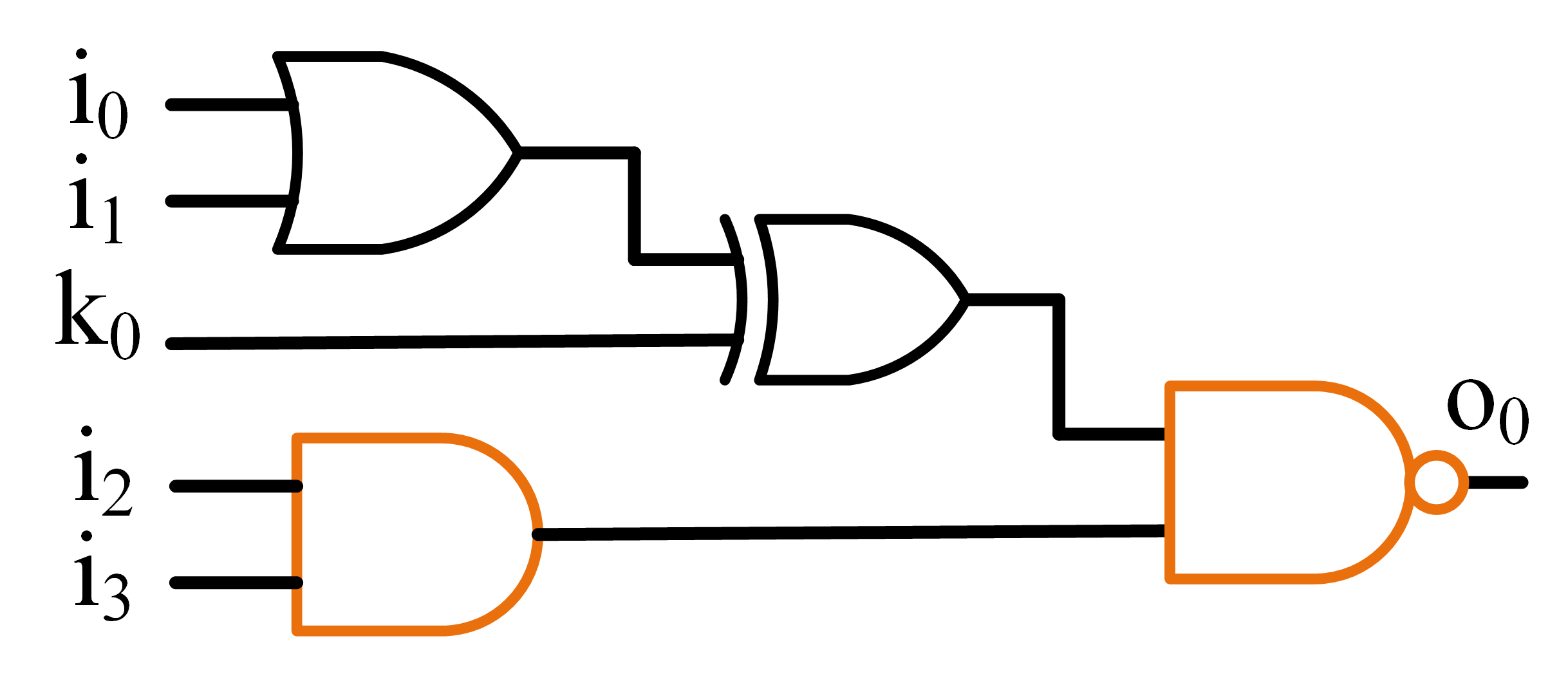}  
\captionof{figure}{Counterexample for the model prediction accuracy of state-of-the-art GNN-based attacks}
\label{fig:counterexample2}
\end{minipage}
\end{figure*}

\subsection{Contributions}\label{Sec:Contribution}
Motivated by the mentioned research gaps, in this paper, we are answering the following questions. First, \textit{why does the accuracy of current GNN attacks differ drastically from the reported key's precision? Can integrating circuit functionality metrics into GNN models result in the discovery of a more relevant key?} We are particularly interested in understanding why hamming distance alone is insufficient to access the effectiveness of GNN attacks and which parameter or set of parameters are meaningful to use here. Second, \textit{what features of the logic-locked circuits led the model to infer the reported key? What is the degree of each feature's influence?} Answering these questions will allow the designer to more securely and confidently pick from many proposed logic locking methods based on the design's trade-offs and circuit characteristics.

Specifically, we introduce a corruptibility-aware and explainable GNN-based OL attack to predict the correct key of logic-locked ICs with a high key precision. To comprehend data behavior and identify relationships between nodes, edges, and node features, an explainable GNN would employ techniques including rule extraction and explainable reasoning \cite{Jegelka:GNN} which, in the context of logic locking attacks, would benefit us in figuring out crucial patterns and topologies that are effective in choosing the key-bit values. The contributions of this paper are threefold:
\begin{itemize}
\item[$\bullet$] Proposing a novel and effective GNN-based OL attack on logic locking that takes the circuit's functionality into account in addition to its structure.
\item[$\bullet$] Providing explainability of the inferred key by the proposed attack that functions as a rule-of-thumb for designers on how to safeguard their precious hardware designs.
\item[$\bullet$] Showcasing the model's prediction accuracy and key precision on seen and unseen logic-locked benchmarks.
\end{itemize}

\subsection{Background}\label{Sec:Background} 
In {\sc xor}-based logic locking \cite{Koushanfar:Logic-locking}, the key-bits can be matched with a combination of random inverters and buffers. Then, key-bit-controlled {\sc xor} gates are used to replace selected buffers and inverters. If an {\sc xor} gate is hiding a buffer, the correct key-bit is ``0'' while if it is hiding an inverter, the correct key-bit is ``1''. Additionally, {\sc mux}-based logic locking \cite{Rajendran:toc13xor} chooses random signals and replaces them with 2-1 {\sc mux}s whose inputs are real signals and random dummy ones, and selectors are the key-bits. As a result, the correct key must select the real signal terminal and avoid the dummy one. Moreover, {\sc lut}-based logic locking \cite{Baumgarten:dtc10lut} is being implemented to IC prefabrication to separate the inputs from the outputs so that a barrier stays between every path from inputs to outputs. In this locking method, the values stored in the Look Up Tables (LUTs) are the key inputs.

While traditional logic locking methods have been attacked by the OG SAT-based attack \cite{Subramanyan:SAT}, post-SAT locking schemes such as SAR-Lock \cite{Rajendran:SAR-Lock} and Anti-SAT \cite{Srivastava:Anti-SAT} have been proposed to exponentially increase the number of input patterns that are required to prune wrong keys by the SAT-based attack. As an advanced post-SAT method, Bilateral Logic Encryption (BLE) \cite{Rezaei:BLE} uses obfuscation and integrated locking on a sensitive component of a circuit. With this approach, the performance overhead is lower than locking the entire circuit, but the security impact, including structural complexity and logic complexity, is transferred to the whole circuit.

Recent studies have shown promising results in the advancement of ML-based OL attacks. SnapShot \cite{Sisejkovic:Snapshot} employs neuro-evolutionary and deep learning methods and is the first of its kind to directly predict the key value from a locked synthesized gate-level netlist. In addition, SAIL \cite{Bhunia:SAIL} recovers the design of a locked circuit in gate-level netlist and extracts circuit features with the help of ML-based structural analysis. While SAIL works mostly on the {\sc xor}-based locked circuits, CutSAIL \cite{Shamsi:CutSAIL} derives missing $k$-cuts from the neighboring logic and predicts the functionality of the missing parts of the locked circuit. However, UNSAIL \cite{Sinanoglu:UNSAIL} inserts unsuited data during the training stage of an ML attack, causing the ML model to predict labels wrong.

More recently, OMLA \cite{Sinanoglu:OMLA} employs a GNN model to predict the keys of a locked circuit by deriving a small subgraph for each key gate. As a result, the key-bit value of a subgraph is also considered its label. GNNUnlock \cite{Sinanoglu:GNNUnlock+}, utilizes GNN for node classification of circuits. The dataset for GNNUnlock is multiple logic-locked circuits of a single benchmark with different key sizes. The model uses adjacency matrices corresponding to the circuit, in which edges and nodes represent wires and gates, respectively. Finally, by mapping the key extraction process to a link prediction problem, UNTANGLE \cite{Sinanoglu:Untangle} gathers concealed links in the lock blocks and then learns the circuit structure, gate features, and link features.

The inability of ML models to be interpreted is one of their main drawbacks. This limitation can be overcome by creating post-hoc explanation procedures for predictions, giving rise to the explainability field \cite{Hlupic:XAI}. For the purpose of producing accurate similarity estimations and discriminative feature representations, SGGNN \cite{Wang:SGGNN} uses graph computation during both the training and testing phases of deep networks. In addition, PGExplainer \cite{Zhang:PGExplainer} uses the trained GNN model as input and offers coherent justifications for the model's predictions. An interesting fact about PGExplainer is that it can be used in an inductive scenario to infer explanations of ambiguous nodes without having to retrain the explanation model.

\begin{figure*}[!t]
    \centering
    \includegraphics[width=\textwidth]{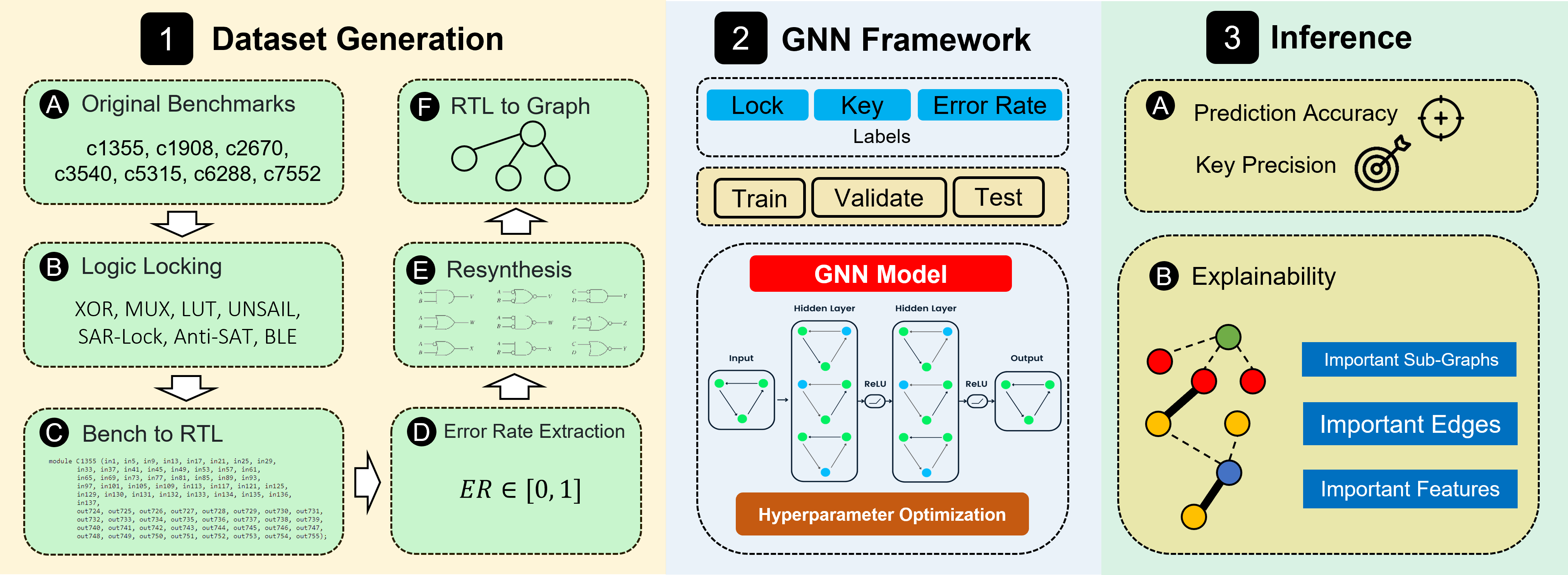}
    \caption{LIPSTICK attack framework}
    \label{fig:BigPicture}
\end{figure*}

\section{Preliminary Study}\label{Sec:Preliminary} 
We want to first make a distinction between two terms: One is ML model \textit{prediction accuracy} which resembles how well a given prediction matches its actual value, and the other is \textit{key precision} which shows how closely a logic-locked circuit under a given key operates to the original circuit.

Now, considering $n$, $m$, and $p$ be the sizes of the input, output, and key respectively, we define the original circuit as $\mathbb{F}:{\{0,1\}}^n \rightarrow {\{0,1\}}^m$, and the locked circuit as $\mathbb{G}:{\{0,1\}}^p \times {\{0,1\}}^n \rightarrow {\{0,1\}}^m$ in which there is a p-bit correct key $K^*=(k_{0}^*, k_{1}^*,..., k_{p-1}^*):{\{0,1\}}^p$ such that $\mathbb{F}(X) = \mathbb{G}(X,K^*)$. We also define the key error rate $ER(K)$ for a given key $K$ of the locked circuit $\mathbb{G}$, as the number of input patterns in which $\mathbb{F}(X) \neq \mathbb{G}(X,K)$, divided by all the input patterns. It is clear that for the correct key, $ER(K^*)=0$. Thus, the $ER$ value is fundamentally dependent on the circuit's functionality rather than its topology.

Let $K^a=(k_{0}^a, k_{1}^a,..., k_{p-1}^a)$ be a p-bit reported key by the attack, the Hamming Distance (HD) of $K^a$ and $K^*$ can be defined as the sum of the bitwise {\sc xor} of the two keys as follows: 
 \begin{equation}
 HD(K^a,K^*) = \sum_{i=0}^{p-1} k_{i}^a \oplus k_{i}^* :{\{0,1,...,p\}}
\end{equation}
 
\textbf{Proposition 1:} The smaller the $HD(K^a,K^*)$, the higher the key precision of the locked circuit $\mathbb{G}$ under $K^a$.

\textbf{Counterexample 1:} Consider the locked circuit in Fig. \ref{fig:counterexample1} with a key size of $p-1$. We increase the key size to $p$ by {\sc xor}ing one of the outputs with additional key-bit $k_{p-1}$. In this case, there is a key $K^a$ in which just the key-bit $k_{p-1}$ is incorrect and all the other key-bit values are the same as $K^*$. In other words, while the HD of $K^a$ is very low (i.e., $HD(K^a,K^*)=1$), the locked circuit $\mathbb{G}$ under $K^a$ outputs differently than the original circuit $\mathbb{F}$ in $100\%$ of the input patterns (i.e., $ER(K^a)=1$).

Hence, Proposition 1 is incorrect since it is possible to design a locking method in which incorrect keys with a small hamming distance from the correct key have high output corruptibility. We can conclude that a key of the locked circuit with a small hamming distance to the correct key does not necessarily outperform a random key with a large hamming distance. To address this issue, we believe that incorporating $ER$ in the training dataset can be useful.

Although, graph representation preserves the topology of the circuit, using an undirected graph for netlist representation is one of the shortcomings of GNNs because the inputs/outputs neighborhood of the netlist will be indistinguishable. Traditional GNN-based attacks cannot effectively distinguish the difference between {\sc xor} and {\sc xnor} key gates due to not taking the circuit functionality into account.

\textbf{Proposition 2:} GNN-based attacks can report an approximate key $K^a$ of the locked circuit $\mathbb{G}$ in which $HD(K^a,K^*)$ is very small.

\textbf{Counterexample 2:} We consider OMLA \cite{Sinanoglu:OMLA} as one of the GNN-based attacks, in which its prediction accuracy has been shown to be on average 80\%. It means, for a reported key $K^a$, it is expected to predict almost 80\% of the key-bits correctly (i.e., $HD(K^a,K^*)=0.2p$). If we replace all the {\sc xor} gates with {\sc xnor} in the benchmarks with {\sc xor}-based locking \cite{Koushanfar:Logic-locking} and push the inverters to the fanouts using bubble pushing, the new correct key will be the complement of the previous one. However, the attack prediction accuracy drops significantly to an average of 56\% (i.e., $HD(K^a,K^*)=0.44p$) which is not much better than reporting a random key.
Fig. \ref{fig:counterexample2} depicts an example of such a transformation on one key-bit. 

The above counterexample shows that Proposition 2 is incorrect and that GNN models are highly dependent on specific gates in the circuit but not on their functional dependencies with each other, and the model prediction accuracy can drop significantly by inverting the key-bits and bubble pushing.

\section{LIPSTICK Attack}\label{Sec:Main}
In this section, we propose \textbf{LIPSTICK}, a corruptibi\textbf{LI}ty-aware and ex\textbf{P}lainable GNN-based oracle-les\textbf{S} a\textbf{T}tack on log\textbf{I}c lo\textbf{CK}ing shown in Fig. \ref{fig:BigPicture}. 

\circleNumber{1}We use seven of the ISCAS'85 \cite{ISCAS85} benchmarks shown in Table \ref{tab:ISCAS} and lock each of them with seven logic locking methods, including {\sc xor}-based locking \cite{Koushanfar:Logic-locking}, {\sc mux}-based locking \cite{Rajendran:toc13xor}, {\sc lut}-based locking \cite{Baumgarten:dtc10lut}, SAR-Lock \cite{Rajendran:SAR-Lock}, Anti-SAT \cite{Srivastava:Anti-SAT}, BLE \cite{Rezaei:BLE}, and UNSAIL \cite{Sinanoglu:UNSAIL} all with a 64-bit key size.

Then we convert .BENCH files of both the original and locked benchmarks into .V using ABC tool \cite{ABC}, and use ModelSim to simulate and extract the $ER$ of 10 random wrong keys ranging from $0$ to $1$ in addition to the correct key. Finally, we apply bubble-pushing to create 10 resynthesized versions of each benchmark.

Overall, our dataset consists of 5,390 elements of data with multiple labels, including a label for the locking method, a label for the designated key for each benchmark (be it correct or wrong), and a label for the $ER$ of that key.

\begin{table}[!t]
\centering
\caption{ISCAS '85 benchmarks  \cite{ISCAS85} information} \label{tab:ISCAS}
{\renewcommand{\arraystretch}{1}%
\begin{tabular}{ c|c|c} 
        \hline
        Bench. & Gates & Functionality \\
        \hline
        \hline
        c1355 & 1503 & 32-bit single-error corrector \\
        \hline
        c1908 & 1289 & 16-bit single-error corrector and double-error detector \\
        \hline
        c2670 & 1262 & 12-bit arithmetic logic unit and controller \\
        \hline
        c3540 & 1403 & 8-bit arithmetic logic unit \\
        \hline
        c5315 & 1350 & 9-bit arithmetic logic unit \\
        \hline
        c6288 & 4703 & 16x16 multiplier \\
        \hline
        c7552 & 1241 & 32-bit adder and comparator \\
        \hline
\end{tabular}
}
\end{table}

\circleNumber{2}We can define GNN as an undirected graph $G=(\mathcal{V},\mathcal{E},\mathnormal{X},\mathnormal{A})$ where $\mathcal{V}$ represents the vertex set, $\mathcal{E}$ is the edge set, $\mathnormal{X}$ constitutes the node feature matrix, and $\mathnormal{A}$ indicates the adjacency matrix of the graph. GNN may learn the embedding of a single node or the complete graph by using the graph's structure and node attributes. In order to compute the intended results, the GNN model adopts neighborhood aggregation by iteratively updating a node's embedding depending on the embeddings of its neighbors. The following equations show the GNN's $i$-th layer, where ${h_v}^{(i)}$ represents the embedding of node $v$ at the $i$-th layer, and and $\mathcal{N}(v)$ shows a set of nodes adjacent to \textit{v}. We use the same method as the Graph Isomorphism Network (GIN) architecture \cite{Jegelka:GNN} to initialize the parameters and consider ${h_v}^{(0)} = X_v$.
\begin{equation}
 {a_v}^{(i)} = AGGREGATE^{(i)} ({{h_u}^{(i-1)}:u \in \mathcal{N}(v)})
\end{equation}
\begin{equation}
 {h_v}^{(i)} = COMBINE^{(i)} ({h_v}^{(i-1)} , {a_v}^{(i)})
\end{equation}

The choice of the GNN method defines which \textit{aggregate} and \textit{combine} functions to use. In this work, the target is graph classification, so for the aggregation function, we use the \textit{readout} function described as follows:
\begin{equation}
h_G = READOUT(\{{h_v}^{(I)} | v \in G \})
\end{equation}
Where $h_G$ is the representation of the entire graph, and the \textit{readout} function acquires the entire graph representation by aggregating node features from the final iteration.

We use the netlist-to-subgraph tool available in \cite{Sinanoglu:OMLA} to extract graphs and subgraphs from .V files of the dataset. The main focus of the training phase is to increase the model's prediction accuracy as well as key precision so that in the validation phase, it predicts a more accurate key with low $ER$. A well-trained model will also provide more meaningful information when fed into a graph explainer tool. A model learns graph features by incorporating a hyperparameter called learning rate, whose exact value is tricky to determine. One naive way is to set it to a constant value, which could either drastically increase the model's training time if the value is too small or stop the model from learning valuable features if it is too large. To provide a reasonable trade-off, we define the learning rate so that after 100 epochs, it gets its 0.01 value for the next 100 epochs. Moreover, we utilize Leaky ReLU as the GNN's activation function to keep the value of $x$ using the maximum function $f(x) = max(0.01x,x)$. Finally, we employ an early stopping strategy to cease training the model if, after five consecutive iterations, the model did not achieve greater accuracy than prior iterations or if the loss value increased to $1$ in order to prevent overtraining the model. At this point, if the model accuracy is still insufficient, we change the sliding window size, pooling window size, and number of layers in the model.

\circleNumber{3}In the post-training phase, we validate the model's prediction accuracy and reported key precision using seen and unseen locked benchmarks. After making sure the model's prediction accuracy and reported key precision are acceptable, we feed the trained model to PGExplainer \cite{Zhang:PGExplainer}. While feature explanation in GNNs is comparable to that in non-graph neural networks, PGExplainer concentrates on explaining graph structures. In order to offer explanations for numerous occurrences, PGExplainer makes use of a parametric explanation network built on a graph-generative model to provide topological explanations.

\begin{table}[!t]
\centering
\caption{OMLA's \cite{Sinanoglu:OMLA} prediction accuracy and reported key precision under different feature maps} 
\label{tab:OMLA}
{\renewcommand{\arraystretch}{1}%
\begin{tabular}{ c|c|c|c} 
        \hline
        Prediction & Key & \multirow{ 2}{*}{Epoch} & \multirow{2}{*}{Feature Map Description} \\ 
        Accuracy  & Precision & &  \\ 
        \hline
        \hline
        80.78 \% & 59.75\% & 350 & Default\\
        \hline
        80.63 \% & 61.33\% & 350 & Random  Assignment\\
        \hline
          
        77.63 \% & 62.29\% & 350 & Highest Assign. to Lowest \#Gates \\
        \hline
         \end{tabular}
    }
\end{table}

\section{Experimental Results}\label{Sec:Experiments}
We implemented \textit{LIPSTICK} on an Intel Core i7-10750H CPU, with a RAM size of 16 GB. 

\subsection{Attack Results}
To compare model prediction accuracy and reported key precision (i.e., (1-$ER$)$\times$100) in state-of-the-art works, we chose OMLA \cite{Sinanoglu:OMLA}, and included different features in the circuit comprehension shown in Table \ref{tab:OMLA}. It is evident that in OMLA, the model's prediction accuracy does not correlate with the reported key precision, and a model's accuracy of 80\% does not assure high key precision. This is because state-of-the-art GNN models focus solely on the structures of the circuits, not their functionality. Another interesting observation here is that OMLA's model prediction accuracy stayed roughly the same when using random feature map assignment compared with the default case, which indicates that the model does not distinguish the gates in its inference.

Table \ref{tab:accuracy} shows \textit{LIPSTICK}'s prediction accuracy and reported key precision using various groups of locking schemes and benchmarks. Unlike column ``5 Random'' which only contains locking schemes that were used in the training set meaning {\sc xor}-based locking \cite{Koushanfar:Logic-locking}, {\sc mux}-based locking \cite{Rajendran:toc13xor}, {\sc lut}-based locking \cite{Baumgarten:dtc10lut}, and SAR-Lock \cite{Rajendran:SAR-Lock}, the ``10 Random'' and ``50 Random'' columns contain unseen locking methods such as Anti-SAT \cite{Srivastava:Anti-SAT} and UNSAIL \cite{Sinanoglu:UNSAIL} as well.

\begin{table}[!t]
\centering
\caption{LIPSTICK's prediction accuracy and reported key precision under random seen and unseen benchmarks. Abreviation guide: X={\sc xor}-based locking \cite{Koushanfar:Logic-locking}, M={\sc mux}-based locking \cite{Rajendran:toc13xor}, L={\sc lut}-based locking, \cite{Baumgarten:dtc10lut}, S=SAR-Lock \cite{Rajendran:SAR-Lock}, B=BLE \cite{Rezaei:BLE}. The word ``Random'' refers to random samples of the validation dataset which includes locking schemes from the training dataset as well as two unseen locking methods: Anti-SAT \cite{ Srivastava:Anti-SAT} and UNSAIL \cite{Sinanoglu:UNSAIL}. The values in ``Random'' columns are the average of key precision for the reported keys.} \label{tab:accuracy}

{\renewcommand{\arraystretch}{1}%
\begin{tabular}{ c|c|c|c|c} 
        \hline
        Locking & Prediction & 5 Random & 10 Random & 50 Random \\ 
        Scheme & Accuracy & Key Prec. & Key Prec. & Key Prec. \\ 
        \hline
        \hline
        X & 92.64\%  & 79.84\% & 75.57\% & 74.97\% \\ 
        \hline
        M & 93.11\% & 79.41\% & 75.44\% & 75.66\% \\
        \hline
        L & 92.75\% & 78.57\% & 75.68\% & 75.54\% \\
        \hline
        S & 93.43\% & 79.19\% & 76.21\% & 75.94\% \\
        \hline
        X,M,L & 85.50\% & 74.86\% & 70.63\% & 70.75\% \\
        \hline
        X,L,S & 84.16\% & 74.33\% & 70.58\% & 70.06\% \\
        \hline
        X,M,S & 82.22\% & 75.78\% & 69.16\% & 68.65\% \\
        \hline
        M,L,S & 84.87\% & 75.44\% & 70.33\% & 69.28\% \\
        \hline
        X,M,L,S & 76.95\% & 69.19\% & 65.39\% & 67.03\% \\
        \hline
        X,M,L,S,B & 51.23\% & 50.63\% & 49.97\% & 50.27\% \\
        \hline
        
    \end{tabular}
    }
\end{table}

Each of the locking methods incorporates a unique algorithm to secure the circuit. Hence, because the structures of each of the circuits are different, each locking method offers different patterns to learn. For this reason, we train the GNN with single locking schemes as well as mixing the locking methods in the dataset. We did not include BLE \cite{Rezaei:BLE} in the single-lock-scheme training because it incorporates the same $ER$ for all the wrong keys, and hence does not let the GNN model learn meaningful information.

The first four rows are the results of single-lock-scheme training, whose prediction accuracy is above 92\% and the average key precision is above 75\%. Comparing Tables \ref{tab:OMLA} and \ref{tab:accuracy}, not only \textit{LIPSTICK} outperforms OMLA in terms of prediction accuracy (i.e., finding a key with low hamming distance to the correct key) but also significantly improves the key precision (i.e., finding a key with low output corruptibility). By including unseen locking methods (i.e., ``10 Random'' and ``50 Random'' columns), key precision accuracy drops a little but is still reasonable due to the fact that these sets include data that GNN was not trained on and did not learn their features.

The results in the final two rows show that, despite the fact that our objective is to provide a universal model that can perform well on various locking methods and different circuit structures, a more diverse dataset (i.e., more than three logic locking methods) does not always result in high prediction accuracy or high key precision. Another reason is that including BLE \cite{Rezaei:BLE} in the training dataset may mislead the model since in BLE any approximate key has high $ER$ and thus low key precision.

\begin{figure}[!t]
    \centering
    \subfloat[] {\includegraphics[width=0.22\columnwidth]{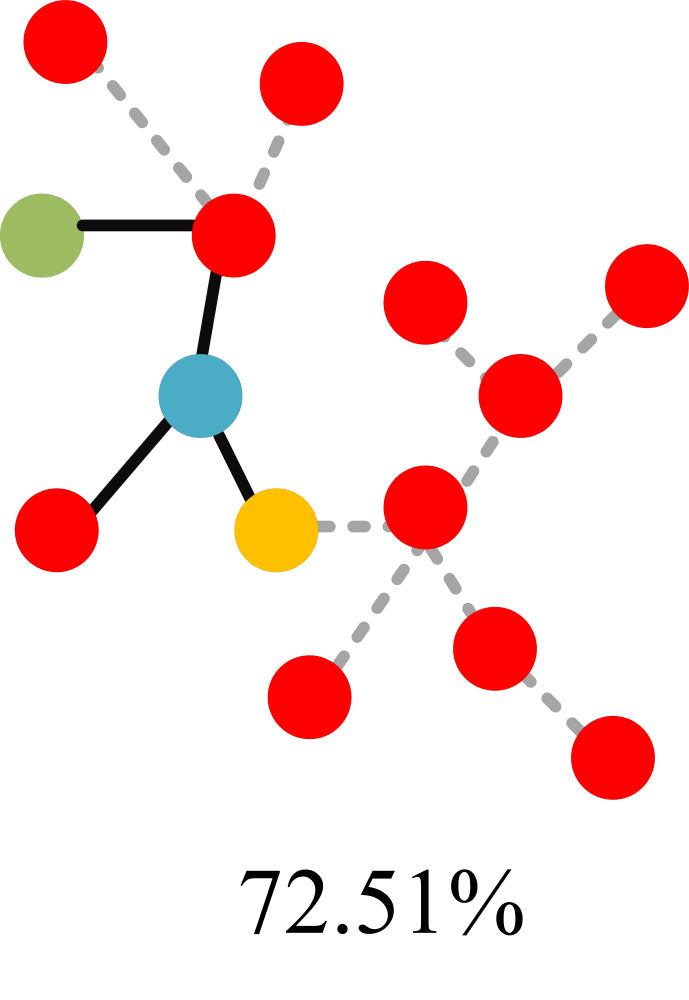}} \hspace{0.1em} 
    \subfloat[]{\includegraphics[width=0.24\columnwidth]{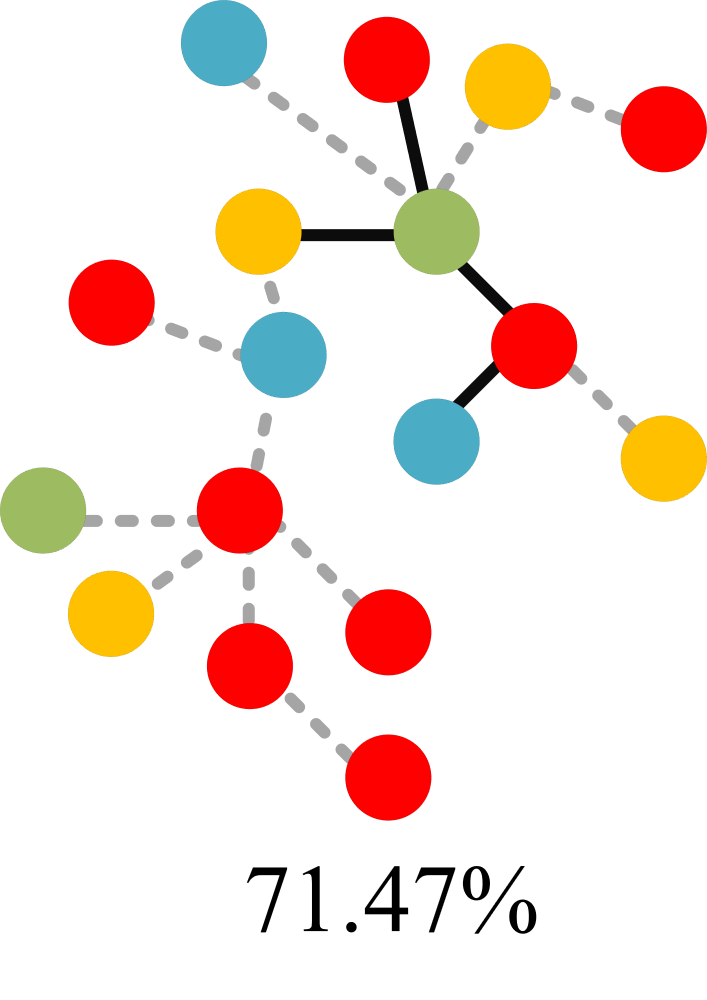}} \hspace{0.5em}
    \subfloat[]{\includegraphics[width=0.2\columnwidth]{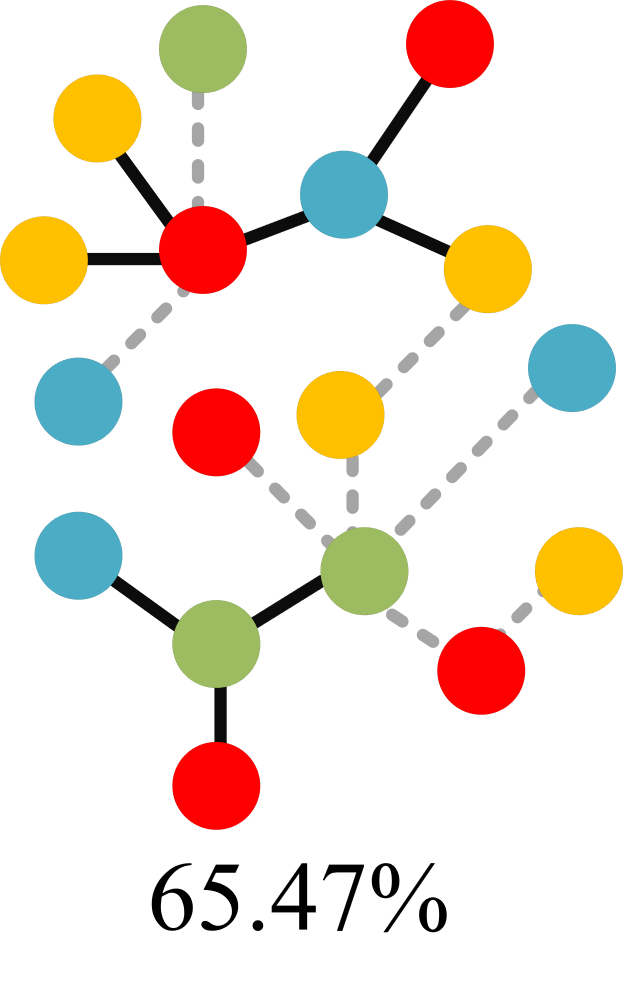}} \hspace{0.5em}
    \subfloat[]   {\includegraphics[width=0.2\columnwidth]{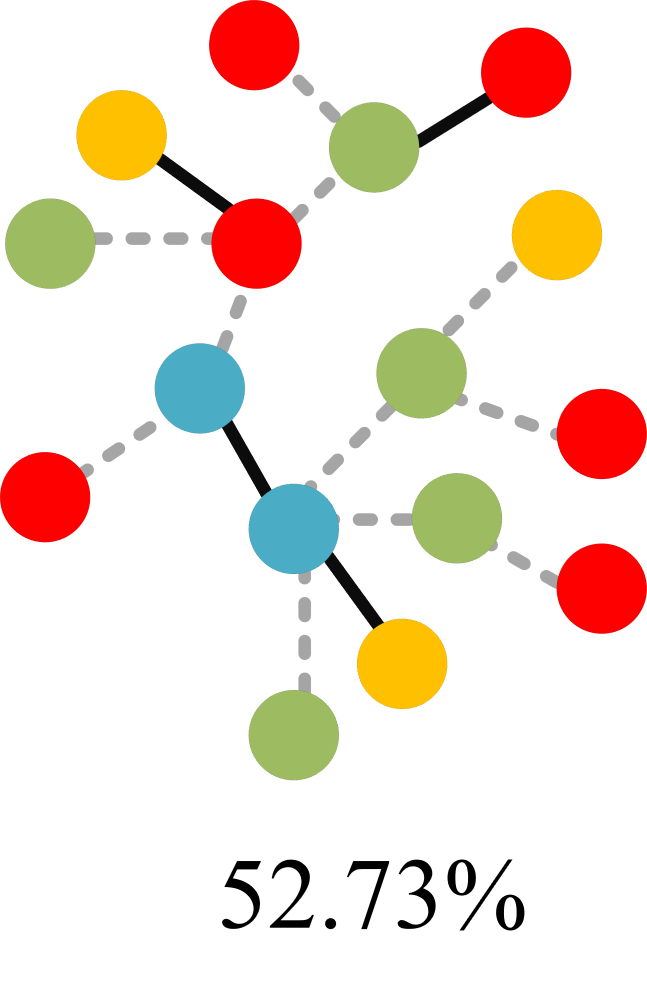}} 
    \caption{Explanation accuracy of LIPSTICK on locking schemes. Bold edges demonstrate a sample pattern that PGExplainer \cite{Zhang:PGExplainer} was able to find. (a) {\sc xor}-based locking (b) {\sc mux}-based locking (c) {\sc xor}-based and {\sc mux}-based methods (d) {\sc xor}-based, {\sc mux}-based, {\sc lut}-based, and SAR-Lock methods}
    \label{fig:PGExplainer}
\end{figure}

\subsection{Explainability Results}
The prediction results of PGExplainer are shown in Fig. \ref{fig:PGExplainer}. In each sub-figure, a sample pattern of the explainable graph prediction is demonstrated. Colored nodes represent different features, and black edges with the corresponding nodes illustrate the patterns that PGExplainer was able to find. Fig. \ref{fig:PGExplainer}a and \ref{fig:PGExplainer}b are the explanation accuracies for the trained graphs with resynthesized benchmarks locked with {\sc xor}-based locking and {\sc mux}-based locking, respectively. By using the default setup of the PGExplainer we achieved 72\% explanation accuracy. While our dataset includes resynthetized versions of the same function, which means that there are multiple structures whose $ER$ are the same under a specific key, PGExplainer is based on one structure per label. In other words, by including $ER$ in the training process, our model goes beyond the structure analysis of graphs, which puts emphasis on the need for the development of an appropriate GNN explanation tool that also goes beyond the structure. Besides that, by providing details on manifold connections and different patterns, PGExplainer gives us information on how to assign features to the input graphs before feeding them to the GNN model. This information is beneficial in allocating rational features to the elements of the input graphs.

Moreover, the explanation accuracy drops in \ref{fig:PGExplainer}c which uses a trained graph as input that was trained with two different logic locking methods. The reason for this accuracy drop is that the PGExplainer should focus on finding more explainable features that do not have common characteristics. Finally, Fig. \ref{fig:PGExplainer}d shows the explanation accuracy result for the trained graph with four locking methods in which the explainer is acting as a random predictor and cannot grasp enough information to provide plausible explanations. It is, however, justifiable by taking another look at the characteristics that each of the locking schemes provides to a circuit. Since the features are diverse, in the training phase, the GNN model did not have enough layers to learn all the various features offered by different locking methods.

\section{Conclusion}\label{Sec:Conclusion}
In this work, we proposed \textit{LIPSTICK}, a corruptibility-aware and explainable GNN-based OL attack on different logic locking methods. \textit{LIPSTICK} incorporates circuit functionality labels in addition to structural parameters into the GNN model with the goal of guiding the model into reporting a more relevant key. In addition, it includes different resynthesized versions of the same circuit, so the model can learn features from different structural views. Furthermore, it involves different logic-locked circuits with both correct and wrong key labels to let the model learn from wrong key insertion too. The experimental results depicted that \textit{LIPSTICK} can achieve both higher model prediction accuracy and higher reported key precision compared to state-of-the-art GNN-based attacks. Moreover, by feeding the trained graphs to a graph explainer tool, we can receive information on how the GNN model is working on the dataset, what the important patterns are, and which components are more important when assigning features to the dataset.

\section*{Acknowledgment}
This work is supported by the National Science Foundation under Award No. 2245247.

\end{document}